\documentstyle[11pt,moriond,epsfig]{article}

\begin{document}
\vspace*{4cm}
\title{THE PARKES MULTIBEAM PULSAR SURVEY}

\author{ I. H. STAIRS, A. G. LYNE, F. CAMILO, N. P. F. MCKAY, D. C. SHEPPARD, D. J. MORRIS }

\address{University of Manchester, Nuffield Radio Astronomy Laboratories,\\
Jodrell Bank, Macclesfield, Cheshire, SK11 9DL, UK}

\author{R. N. MANCHESTER, J. F. BELL}

\address{Australia Telescope National Facility, CSIRO, P.O.Box 76,\\ Epping
NSW 2121, Australia}

\author{ V. M. KASPI, F. CRAWFORD}

\address{Massachusetts Institute of Technology, Center
for Space Research, \\70 Vassar Street, Cambridge, MA 02139, USA}

\author{N. D'AMICO}

\address{Osservatorio Astronomico di Bologna, via Ranzani 1, 40127
Bologna, Italy}

\maketitle\abstracts{ 
The Parkes multibeam pulsar survey is a high-frequency, fast-sampled
survey of the Galactic Plane, expected to discover at least 500 new
pulsars.  To date, over 200 pulsars have been found, including several
young pulsars and at least one with a very high magnetic field.  Seven
of the new stars are in binary systems; this number includes one
probable double-neutron-star system, and one pulsar with an extremely
massive companion.}

\section{Introduction and Survey Parameters}

Large-area searches for new pulsars require multiple filtered channels
across a wide bandwidth, in order to search a wide range of dispersion
measures (DMs), and rapid sampling, in order to be sensitive to
fast-spinning pulsars.  Pulsars are steep-spectrum objects; therefore
historically most pulsar searches have been conducted at
frequencies of a few hundred MHz.  However, this choice of frequency
introduces certain undesirable selection effects.  In particular,
pulse broadening due to dispersion smearing and multipath interstellar
scattering limits the sensitivity of the search to distant (high-DM),
short period pulsars.  Also, the high galactic background temperature
significantly increases the telescope system noise near the galactic
plane; thus a higher observing frequency, such as 1.4\,GHz, is a
better choice for surveys concentrating on this region of the sky.

Two such high-frequency surveys have been conducted in the recent
past: one used the 76-m Lovell telescope at Jodrell Bank in the
U.K. to search the Northern part of the Galactic plane, finding 40
pulsars,\cite{clj+92} while a complementary search at the 64-m Parkes
telescope in Australia found 46 pulsars.\cite{jlm+92}

Recently, a new multibeam receiver\,\cite{swb+96} has been installed
at Parkes, allowing simultaneous observations of 13 patches of the
sky.  The system was originally designed for a survey of HI in the
local universe, but is also being used for a pulsar search in a
10$^\circ$-wide strip along the galactic plane.  With the multibeam
receiver, the time required to survey a given part of the sky is
reduced by a factor of 13.  Moreover, in the time since since the
completion of the earlier surveys, receiver sensitivities have
increased, and computer and recording-medium speeds have improved
greatly, allowing faster sampling of wider bandwidths, and longer
integration times.  The new survey, and its planned counterpart at
Jodrell Bank, make full use of these advancements, and will therefore
be considerably more sensitive to fast and distant pulsars than the
earlier searches.  A comparison of the parameters of the four surveys
is given in Table~\ref{tab:surveys}.

\begin{table}[t]
\caption{Parameters of Four 21-cm Pulsar Surveys\label{tab:surveys}}
\vspace{0.4cm}
\begin{center}
\begin{tabular}{|l|l|l|l|l|}
\hline
                & Jodrell\,\cite{clj+92} & Parkes\,\cite{jlm+92} & Parkes  & Jodrell  \\
\hline
& & & & \\
Nbeams     &    1    &    1     &   13    &    4   \\
$|b|$           & $<1^\circ$ & $<4^\circ$  & $<5^\circ$ & $<5^\circ$\\
$l$       & $-5^\circ$--$100^\circ$    & $-90^\circ$--$20^\circ$    
& $-100^\circ$--$50^\circ$  & $50^\circ$--120$^\circ$  \\
$t_{int}$ (min) &   10    & 2.5      &   35    &   35   \\
$t_{samp}$ (ms) &   2.0   & 1.2 &  0.25   &  0.25  \\
Bandwidth (MHz) & $2\times 8\times5$   &$2\times 64\times5$   &
$2\times 96\times3$    &  $2 \times 32\times3$    \\
$S_{sys}$ (Jy)  &   60    & 70       &  36     &  30    \\
$S_{min}$ (mJy) &   1.2   & 1.0     &  0.15   &  0.2   
\\
Pulsars found  &   40    & 46       & in progress  & forthcoming   \\
Acceleration search & No  & No       & Yes     & Yes    \\
\hline
\end{tabular}
\end{center}
\end{table}

\begin{figure}[t]
\begin{center}
\psfig{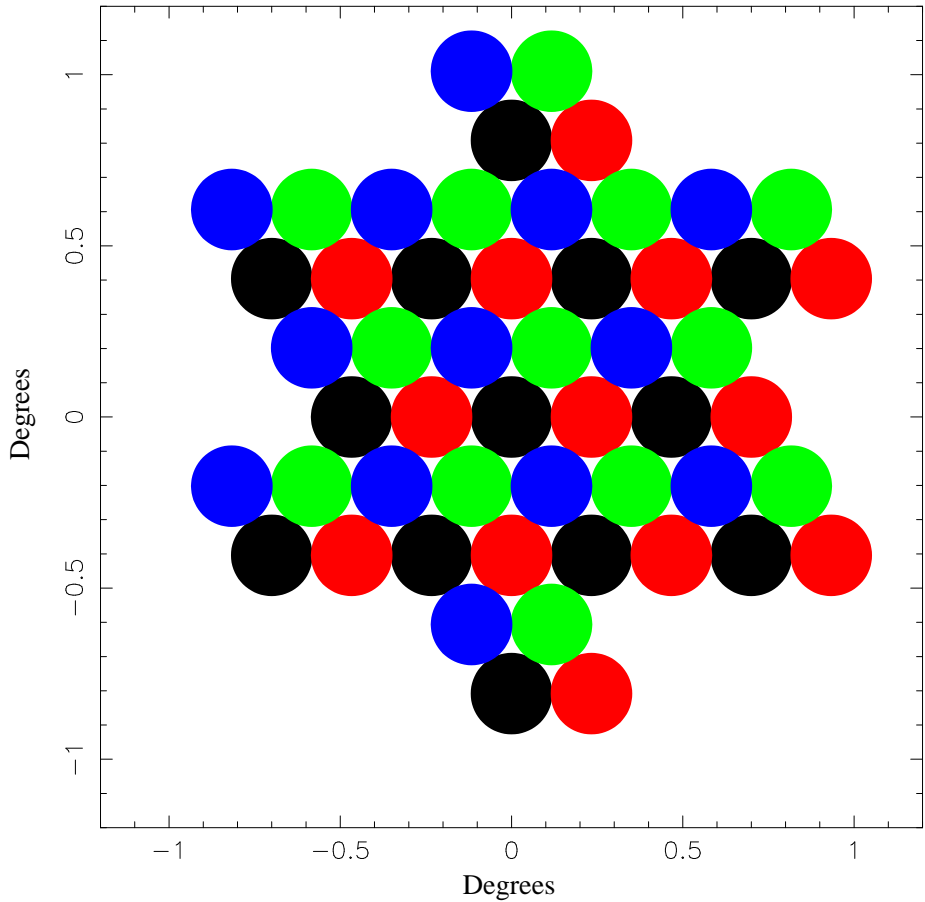}
\caption{Four-pointing tessellation pattern of the 13 beams on the sky.
\label{fig:tess}}
\end{center}
\end{figure}

Each beam of the new receiver is approximately 0.25$^\circ$ wide; the
beams centres are spaced 2 beamwidths apart.  The 35-min pointings are
arranged in groups of four, as shown in Figure~\ref{fig:tess}, giving
complete sky coverage on a hexagonal grid.  A total of 2670 pointings
of the 13 beams are required to cover the survey region. 

Analysis procedures are similar to those used in previous
surveys,\cite{mld+96} with dedispersion followed by transformation
into the modulation frequency domain using a Fast Fourier Transform
(FFT) and harmonic summing to optimize sensitivity to the typically
short duty-cycle pulses.  The pulse frequencies of pulsars in close
binary orbits may be subject to significant changing Doppler shifts
during the 35-minute pointings; hence segmented ``acceleration
searches'' will be conducted to look for drifting periodicities.

\begin{figure}[t]
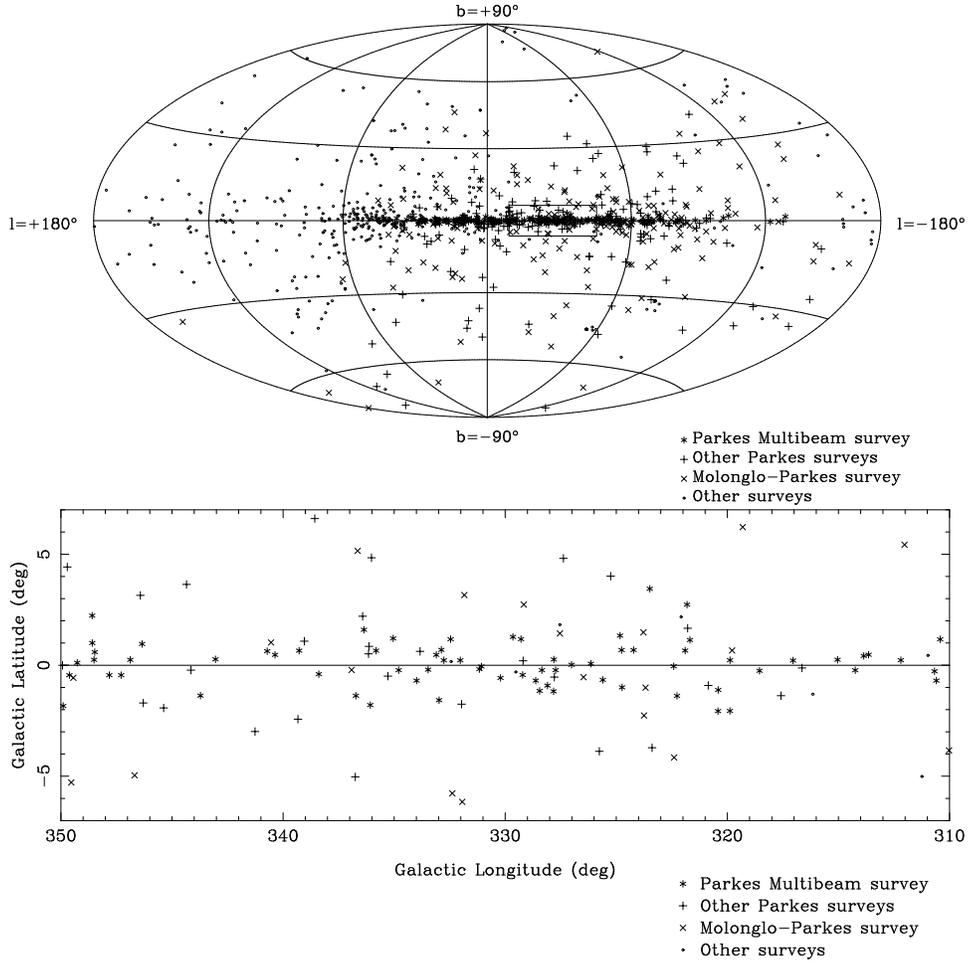

\begin{center}
\psfig{figure=stairsp2.ps,angle=270,width=5in}
\vspace{0.5in}
\psfig{figure=stairsp3.ps,angle=270,width=5in}
\caption{The 220 new Parkes multibeam survey pulsars in galactic coordinates; the lower panel is an enlargement of the boxed region.
\label{fig:gallb}}
\end{center}
\end{figure}

Suspects from the initial processing are scheduled for reobservation
to confirm their identity as pulsars. The centre beam of the
multibeam system is used to make five observations, typically of 6-min
duration, at the nominal position and four surrounding points spaced
$9^\prime$ from the nominal position. In most cases this gives
detections at two or three positions, allowing an improved position to
be determined, as well as confirming the pulsar. Where no detection
results from these observations, a 35-min observation is made at the
nominal position.

After confirming each pulsar, a series of timing observations is begun
at either or both of the Parkes 64-m telescope and the Lovell 76-m
telescope at Jodrell Bank.  Almost all of the detected pulsars north
of declination $-35^\circ$ (some 40\% of the total to date) are being
timed only at Jodrell Bank.

For both Jodrell Bank and Parkes data, each pulse profile obtained by
summing over an observation is convolved with a high signal-to-noise
``standard profile'' for the corresponding pulsar, producing a
topocentric time of arrival (TOA). These are then processed using the
TEMPO program (see WWW address http://pulsar.princeton.edu/tempo)
which converts them to barycentric TOAs at infinite frequency and
performs a multi-parameter fit for the pulsar parameters.  Barycentric
corrections are obtained using the Jet Propulsion Laboratory DE200
solar-system ephemeris.\cite{sta82} Except for especially interesting
cases, it is proposed to make timing observations of each pulsar over
about 12 months, giving an accurate position, period, period
derivative and dispersion measure. This provides the basic parameters
necessary for follow-up studies such as investigations of the Galactic
distribution of pulsars and studies of the interstellar medium. Many
of the pulsars are relatively young and subject to timing
irregularities and initial data on these will also be obtained.

\section{Survey Status and New Discoveries}\label{sec:status}

As of Oct. 1998, approximately 40 per cent of the required pointings
had been observed, and about 40 per cent of these had been
processed. Confirmation observations had been made on most of the
better suspects, resulting in the detection of 220 previously unknown
pulsars, including the 1000th pulsar ever discovered.  More than 100
previously known pulsars had also been detected. These results show
that the survey is going to be outstandingly successful. The discovery
rate is an unprecedented one pulsar per hour of survey observing time,
more than an order of magnitude better than any previous pulsar
survey. 

A plot of the distribution in galactic coordinates of the new pulsars
is presented in Figure~\ref{fig:gallb}, with the boxed region shown in
greater detail.  The great success of the multibeam survey relative to
previous searches is easily seen.

\begin{figure}[t]
\begin{center}
\psfig{figure=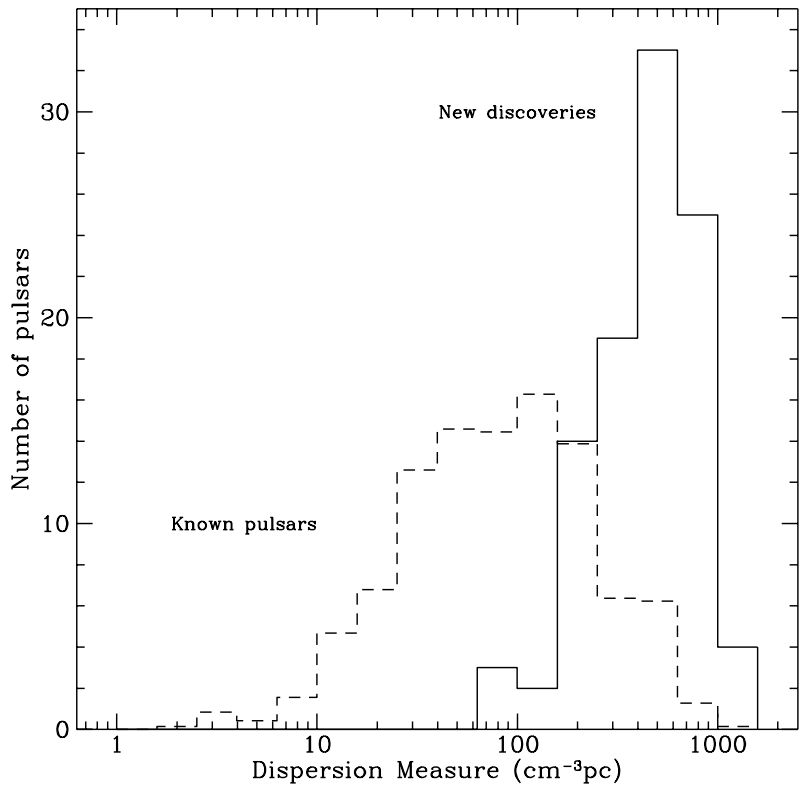,height=4.5in}
\caption{Dispersion measure distribution of all previously-known 
non-globular-cluster pulsars (dashed line) and the new Parkes
multibeam survey pulsars (solid line), with normalized areas.  The
distributions reflect the greater sensitivity of the current survey to
distant pulsars.
\label{fig:dmdist}}
\end{center}
\end{figure}

Due to the choice of high observing frequency, narrow filterbank
channels and fast sampling, the present survey is extremely sensitive
to distant pulsars.  Figure~\ref{fig:dmdist} plots the
dispersion-measure distribution of the newly-found pulsars and of all
previously-known pulsars which are not in globular clusters.  Of the
220 new pulsars discussed above, 175 have DM$\,>\,250$\,pc\,cm$^{-3}$;
this has nearly tripled the previous number of 99 such pulsars.  In
contrast, there are very few new pulsars with
DM$\,<\,100$\,pc\,cm$^{-3}$, as most of this population has been found
in earlier surveys.

The survey has uncovered 8 apparently isolated pulsars with spin
periods between 45\,ms and 90\,ms.  It is probable that these are
young pulsars, though this will only be confirmed by further timing
observations, from which their period derivatives and hence their
characteristic ages ($\tau_c\,=\,P/(2{\dot P})$) may be determined.
There is also a pulsar with a relatively long spin period (0.4\,s),
but an extremely short characteristic age of 1600 years.  The
implication is that this pulsar has an extremely high magnetic field.

Another area of great success for the search has been in the discovery
of binary pulsar systems: 7 have been found to date.  Five of these
pulsars have spin periods between 9\,ms and 88\,ms, and nearly
circular orbits.  These are likely to be neutron-star--white-dwarf
systems, and the Keplerian mass functions indicate that some of the
companions may in fact be heavy CO white dwarfs.  Prior to the start
of this survey, only four such neutron-star--CO-white-dwarf systems
were known;\,\cite{cam96c} the fact that they represent a greater
fraction of the new discoveries indicates that this pairing may not
be as rare as previously believed.
\clearpage

\begin{figure}[t]
\begin{center}
\psfig{figure=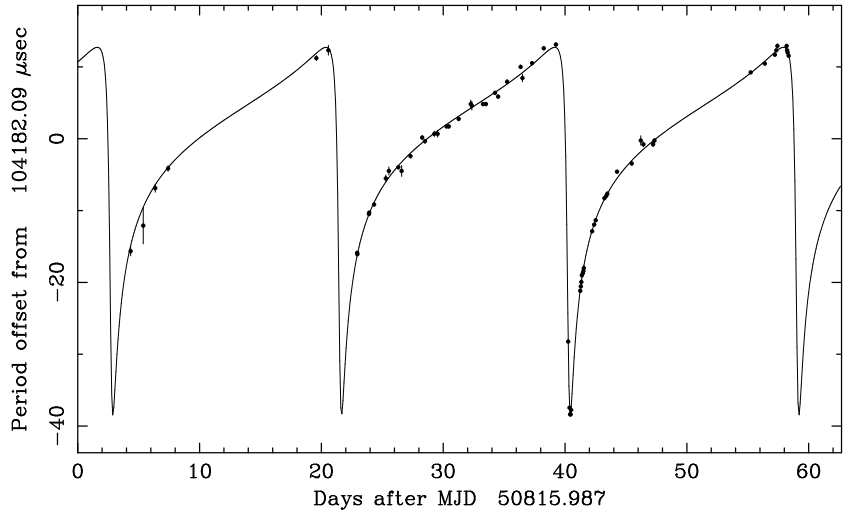,angle=270,width=5in}
\caption{Velocity curve of the probable double-neutron-star binary pulsar, 
across three cycles of the 18-day orbit.
\label{fig:3orbits}}
\end{center}
\end{figure}

The two remaining systems are more unusual.  One has an 18-day orbit
with large eccentricity ($e\simeq0.8$), and a minimum companion mass
of 0.8\,$M_\odot$.  This is almost certainly a double-neutron-star
system, the first to be discovered in the Southern hemisphere.  The
velocity curve for this pulsar across three orbits is displayed in
Figure~\ref{fig:3orbits}.  While this system, like the similar
PSR~J1518+4904,\cite{nst96} is not in a close enough orbit to permit
tests of general relativity, it will nevertheless provide a valuable
contribution to our understanding of the formation and population
statistics of such binaries.  The last system is perhaps the most
intriguing: the pulsar is in an 8-month, highly-eccentric orbit, and
has a companion of minimum mass 11\,$M_\odot$.  The nature of the
companion is as yet undetermined; this pulsar system will therefore be
the target of optical and/or infrared-wavelength observations in the
near future.

\section{Future Prospects}

The Parkes multibeam pulsar survey is proving to be extremely
successful.  Though the discovery rate will decline from the current
one pulsar per hour of observation as regions further from the
Galactic plane are surveyed, predictions suggest that the survey may
detect as many as 500 previously unknown pulsars, providing a large
increment to the total of about 750 pulsars known before the survey
commenced, and a significant database for many different follow-up
studies.  If the fraction of binary pulsars found remains constant, it
is reasonable to expect 15 or 20 new systems, some of which may well
prove even more fascinating than those already discovered.

\section*{Acknowledgments}
The Australia Telescope is funded by the Commonwealth of Australia
for operation as a National Facility managed by CSIRO.

\end{document}